# Binary information propagation in circular magnetic nanodot arrays using strain induced magnetic anisotropy

*M. Salehi-Fashami[1], M. Al-Rashid[2,3], Wei-Yang Sun[4], P. Nordeen[4], S. Bandyopadhyay[3], A.C. Chavez[4], G.P. Carman[4] and J. Atulasimha[2,3]\**

[1]Department of Physics, Univ. of Delaware, Newark, DE 19716.

[2]Department of Mechanical and Nuclear Engineering, Virginia Commonwealth Univ., Richmond, VA 23284.

[3]Department of Electrical and Computer Engineering, Virginia Commonwealth Univ., Richmond, VA 23284.

[4]Department of Mechanical and Aerospace Engineering, Univ. of California, Los Angeles, CA 90095.

ABSTRACT: Nanomagnetic logic has emerged as a potential replacement for traditional CMOS-based logic because of superior energy-efficiency[1,2]. One implementation of nanomagnetic logic employs shape-anisotropic (e.g. elliptical) ferromagnets (with two stable magnetization orientations) as binary switches that rely on dipole-dipole interaction to communicate binary information[2–10]. Normally, circular nanomagnets are incompatible with this approach since they lack distinct stable in-plane magnetization orientations to encode bits. However, circular magnetoelastic nanomagnets can be made bi-stable with a voltage induced anisotropic strain, which provides two significant advantages for nanomagnetic logic applications. First, the shape-anisotropy energy barrier is eliminated which reduces the amount of energy to reorient the dipole. Second, the in-plane size can be reduced (~20nm) which was previously impossible due to thermal stability issues. In circular magnetoelastic nanomagnets, a voltage induced strain stabilizes the magnetization even at this size overcoming the thermal stability issue. In this paper, we analytically demonstrate a binary 'logic wire' implemented with an array of *circular* nanomagnets that are clocked with voltage-induced strain applied by an underlying piezoelectric substrate. This leads to an energy-efficient logic paradigm orders of magnitude superior to existing CMOS-based logic that is *scalable* to dimensions substantially smaller than those for existing nanomagnetic logic approaches. The analytical approach is validated with experimental measurements conducted on dipole coupled Ni nanodots fabricated on a PMN-PT sample.

Strain control of a magnetoelastic materials magnetic state has been explored experimentally in a number of material systems[11–15] but experimental studies on single domain nanomagnets has only arisen recently[16]. In these single domain studies few papers have focused on logic based operations. Several of the theoretical papers have established strain clocked nanomagnetic memory and logic employing single-domain nanomagnets is energy efficient[5–8] and require as little as ~1aJ/bit of energy to switch. Thus, the total energy required to switch a nanomagnet with strain is *2-3 orders* of magnitude less than that required for switching current CMOS devices[17] and *4-5 orders* of magnitude less than for nanomagnets clocked with a current-generated magnetic field[9] or spin transfer torque (STT)[18]. Other energy-efficient methods of clocking nanomagnets include the use of Spin Hall Effect[10], and spin orbit torque[19].

In logic applications, nanomagnets are typically designed with high magnetic anisotropy (either shape or perpendicular magnetic anisotropy) energy barriers ($\Delta U \geq 50 \ k_B T$, where $k_B$ is the Boltzmann constant and $T = 300$ K). This energy barrier is required for two reasons: first, the anisotropy produces two distinct stable magnetization orientations to encode the binary logic bits '0' and '1'. Second, the energy barrier prevents the magnetization from randomly flipping between the two stable states in the presence of thermal noise (the probability of spontaneous flipping is $\sim e^{-\Delta U / k_B T}$). The latter feature expands the usefulness of 'non-volatile' nanomagnetic logic because now the same device can be used as both 'logic' and 'memory'.

However, some nanomagnetic elements can be volatile for a device that implements non-volatile logic. Specifically, only the nanomagnets storing the output bits need to be non-volatile and require a thermal energy barrier ($\Delta U \geq 50 \ k_B T$, where $k_B$ is the Boltzmann constant and T is 300 K) between the degenerate "0" and "1" states at room temperature. The other nanomagnetic elements in the logic devices merely carry out logic operations rather than store bits of information and thus may be volatile. Therefore, these other nanomagnets can be small, super-paramagnetic, and circular. Recent experimental studies have shown that anisotropy created by different methods can transform super-paramagnetic nanoparticles at room temperature to single domain non-volatile ferromagnets[20–22]. Specifically, it has been recently demonstrated that Ni nickel nanoparticles can be switched between a super-paramagnetic state and a single-domain ferromagnetic state at room temperature by application of a voltage induced biaxial strain that changes the magnetic anisotropy[22]. This provides two distinct advantages. First, this lowers the amount of energy required to propagate the information along the chain to the final non-volatile bit. Second, the size of these elements can be made ultra-small to increase processing *density*.

This results in an extremely energy efficient nanomagnetic logic device that improves scalability to smaller feature sizes.

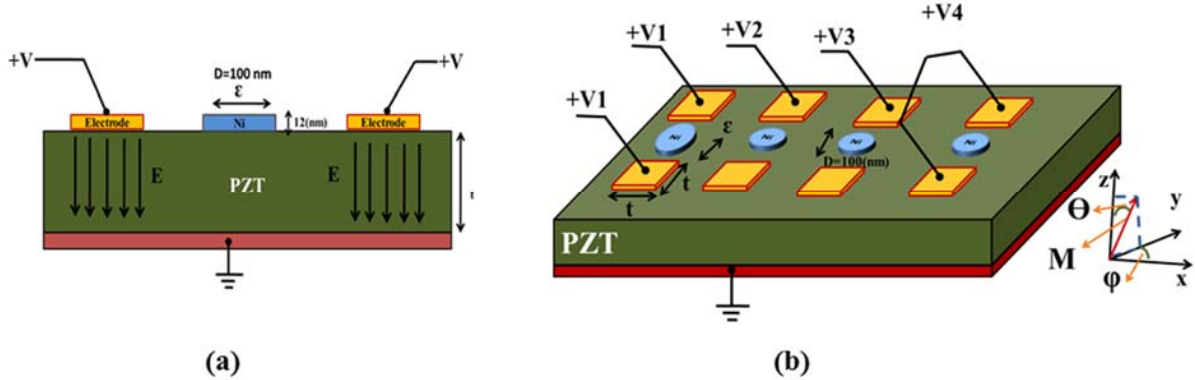

**(a)**                                    **(b)**

**Figure 1.** Implementation of a logic wire. (a) A multiferroic circular nanomagnet with diameter of 100 nm and thickness of 12 nm. Note: A biaxial stress in developed between the electrodes in the manner described in Ref 21 when an electrostatic potential is applied between them and the grounded substrate. (b) Chain of dipole coupled nanomagnets with center to center separation of "d" that can be clocked sequentially using a local clocking scheme that generates local stress only under the selected nanomagnet.

A critical component of the dipole-coupled nanomagnetic logic systems is a "binary wire" that propagates a logic bit unidirectionally from one end of the wire to the other[2-5, 7-10]. In this paper, we theoretically simulate a binary wire implemented with a linear array of dipole-coupled *circular* nanomagnets subjected to room-temperature thermal noise. In this simulation, voltage induced Bennett clocking[23] of the nanomagnets is achieved with strain produced by a piezoelectric thin film deposited onto a silicon substrate as shown in Figure 1. Here voltage is applied to electrode pairs to overcome the substrate clamping issues imposed on a thin film piezoelectric (~500 nm or less) as proposed by Cui et al.[24] to generate a bi-axial strain that is transferred to the nanomagnet. (Supplementary material Section A1 describes the in-plane magnetic anisotropy created by applying a voltage induced strain that turns a circular ferromagnet into a bistable nanomagnet element). By sequentially applying a voltage, propagation of the information encoded in the magnetic moments of the nanomagnetic wire is achieved. Further, the ability to control dipole-dipole interaction is experimentally demonstrated using Ni nanodots of 100 nanometer diameter on a PMN-PT substrate. Test data shows that the dipole coupling can be modulated with an applied electric field. We note that these experiments use *global* clocking as the tests are performed on a bulk PMN-PT substrate rather than thin film PZT. The strain mediated voltage control of magnetic anisotropy in circular nanomagnets may spawn a new genre of efficient nanomagnetic logic hardware implemented with ultra small circular super-paramagnetic structures.

## INFORMATION PROPAGATION IN AN ARRAY OF CIRCULAR NANOMAGNETS WITHOUT SHAPE-ANISOTROPY IN THE PRESENCE OF THERMAL NOISE

Here we study information propagation without stress and then with sequential application of stress to show how the latter in necessary to propagate information in circular nanomagnets in the presence of thermal noise.

### No stress case

In this paper a chain of dipole coupled magnetoelastic nanomagnets are considered as shown in Figure 2a. The initial element is an elliptical nanomagnet representing the encoded input bit followed by circular nanomagnets. The elliptical nanomagnet has two in-plane stable states along the major axis – "up" ($\phi = 90^0$) encoding binary bit '1' and "down" ($\phi = -90^0$) encoding binary bit '0'. When the elliptical nanomanget (element 1) input bit is '1', i.e. the magnetization is pointing "up" ($\phi$ =90º), its magnetic dipole influences the next circular nanomagnet (element 2) to point "down" ($\phi$ = -90⁰). Furthermore, dipole-dipole coupling between element 2 and the next circular nanomagnet (element 3) causes the third element to rotate towards the down ($\phi$ = -90⁰) horizontal. As the number of circular namomagnets increases in the line, the dipole-dipole dictated by the first elliptical nanomagnetic producing the anti-parallel vertical ordering ("up", "down", "up", "down" and so on) gives way to dipole-dipole coupling present between the circular nanomagnetic that dictates a horizontal orientation (parallel alignment) of the magnetization along the axis of the chain of nanomagnets, i.e. $\phi = 0^\circ$). Thus, the 'logic wire' produced by the ellipse eventually fails.

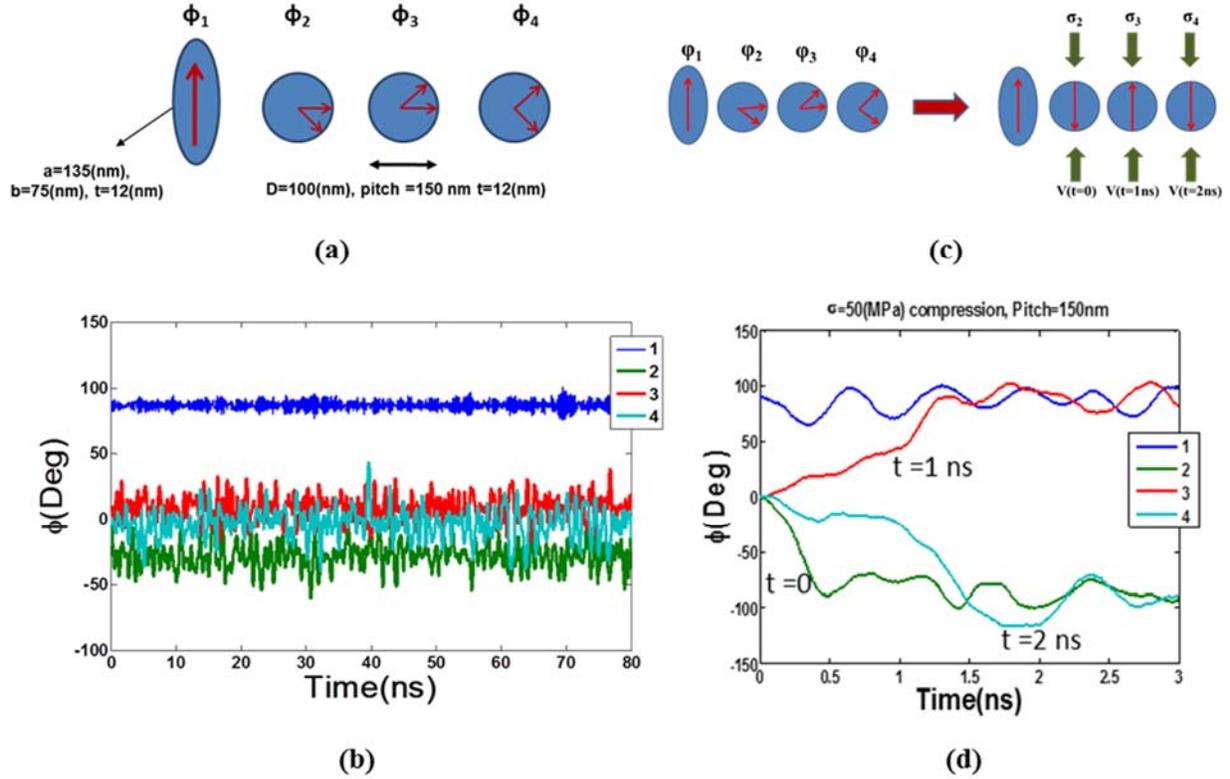

**Figure 2.** Information propagation in a binary wire composed of circular nanomagnets. (a) Schematic view of dipole coupled circular nanomagnets forming a "logic wire" preceded by a high shape-anisotropy nanomagnet acting as the input bit host. (b) Fluctuations of nanomagnet's in-plane magnetization orientation about the mean orientation vs. time in the absence of stress. (c) Sequential clocking of the circular nanomagnets with compressive mechanical stress. (d) In-plane magnetization dynamics of dipole coupled nanomagnets versus time, showing that stress promotes "logic restoration" or near "up" or near "down" orientation of the magnetization in each nanomagnet.

Figure 2b shows the Landau Lifshitz Gilbert (LLG) simulation in the presence of thermal noise results for in-plane magnetization versus time for the configuration illustrated in Figure 2a that represents the base state or magnetization orientation in the absence of a voltage induced strain.

The stochastic magnetization dynamics of the nanomagnets in the array under the influence of dipole fields and stress was simulated using the Landau-Lifshitz-Gilbert (LLG) equation[27]:

$$\frac{d\vec{M}(t)}{dt} = -\gamma \vec{M}(t) \times \vec{H}_{eff}(t) - \frac{\alpha\gamma}{M_s}\left[\vec{M}(t) \times \left(\vec{M}(t) \times \vec{H}_{eff}(t)\right)\right] \tag{1}$$

where $\gamma$ is the gyromagnetic ratio, $\alpha$ is the Gilbert damping constant in the nanomagnet[25] and $\vec{H}_{eff}^i$ is the effective magnetic field acting on the magnetization vector. The effective field incorporates the effect of stress, dipole coupling from neighbors, and random thermal noise[26,27]. The details of the simulation can be found in the supplementary material section A2.

The results show the first elliptical nanomagnet's magnetization oscillates (because of thermal fluctuations) but is stable around $\phi = 90^\circ$. In sharp contrast, the second element (i.e. circular nanomagnet) oscillates around $\phi = -30$ while the third oscillates around $\phi = \sim 10^\circ$. When the fourth element is reached the logic wire produced by the ellipse is non-existent with $\phi = \sim 0^\circ$. The *distributions* of the second, third, and fourth nanomagnet's in-plane magnetizations around their respective means are shown in the supplementary material section A3 in Figure S2 (a, b and c). Clearly, the influence of the input nanomagnet's magnetic state (i.e. elliptical nanomagnetic) decays with distance and is virtually undetectable past the third element. Thus, all information is lost beyond the third element since all subsequent nanomagnets orient their magnetizations along the horizontal.

**Sequential straining of the circular nanomagnets (Bennett clocking) correctly transmits bit information down the logic wire**

Figure 2c shows a schematic representation when a voltage (V) is applied sequentially to generate strain in each of the circular nanomagnets. Prior to t=0, voltage is absent and the first "input" elliptical nanomagnet's magnetization points at 90° (in the upward direction) while the other nanomagnets' magnetization are assumed to point at 0° (to the right). A sequential voltage is applied, starting at t=0 with 1ns delay onto each consecutive nanomagnet starting with the second element. The voltage induces ~250 ppm compressive strain (we assume a very conservative value, instead of ~1000 ppm used in Cui et al.[24] and Wu et al.[28]] that can be generated in each circular nanomagnet producing an anisotropy favoring alignment with the y-axis shown in Figure 1. The voltage induced anisotropy due to a compressive strain is caused by the negative magnetostrictive properties of Ni.

Figure 2d shows the in-plane magnetization dynamics that result from the application of a sequential voltage to the elements. For these results a voltage is applied at t=0 to the second element, at t=1ns to the third element, and at t=2ns to the fourth element. As can be seen, the second nanomagnet in the chain rotates toward -90° after the application of the voltage at t=0 and begins oscillations around -90° after about 0.5 ns. This deterministic rotation, i.e. counter-clockwise as contrasted with clockwise, is caused by the dipole-dipole interaction present with the adjacent elliptical nanomagnet. Here it is important to point out that the motion of this circular nanomagnetic also influences it neighbor, the third element rotates partially between t=0 and t=1ns.  At t=1 ns a voltage is applied to the third element producing a rotation/stabilization toward the 90° direction.  The periodic oscillations in this element about this equilibrium point occurs at approximately t=1.3ns or within 0.3ns of the voltage applied. Here it is again important to point out that the motion of this circular nanomagnet also influences it neighbor, the fourth element which also rotates partially before a voltage is applied to it. The application of this voltage (i.e. t=2 ns) stabilizes the fourth element with an orientation at -90°.  These results show that a chain of many nanomagnets can be clocked sequentially with a voltage to propagate logic along this chain.

The results in Figure 2c demonstrate that a voltage produces sufficient compressive strain to each nanomagnet to significantly lower the energy of the "up" and "down" states ($\phi = 90^0, -90^0$).  This essentially promotes promote anti-ferromagnetic ordering of magnetizations in the "up" and down" states This anti-ferromagnetic orders is the result of inter-magnet dipole coupling and the voltage induced  stress anisotropy that make this configuration energetically more favorable than the magnetizations pointing horizontally. Furthermore, the results show that information can successfully be transferred along a chain of circular nanomagnets at the rate of 1 bit/nanosecond between two neighboring nanomagnets.

The numerical simulations assume two mechanisms dominate the magnetization reorientation. These two effects are (1) dipole coupling between nanomagnets and (2) stress induced magnetic anisotropy through the magnetoelastic effect. To demonstrate that these are the two dominant mechanisms a series of experiments were designed to establish the dipole effect and its relation to stress induced magnetic anisotropies.  These effects can be quantified by measuring the averaged M-H curves of an ensemble of nanomagnets and focusing on a large pitch array, a small pitch array (dipole coupled) without stress and a small pitch array subjected to stress. By measuring the remanence changes along select directions specific conclusions can be drawn about the relative magnitudes and the ability to control dipole coupling in the proposed logic device supporting the analytical conclusions. These results are discussed below.

**EXPERIMENTS**

Two different samples were tested consisting of Nickel circular (100nm) nanodot arrays patterned on a 10mm x 10mm x 0.5mm single crystal (011) PMN-PT substrate sourced from TRS Technologies, Inc., USA. The arrays are 1x1 mm and isolated from each other with a 2mm separation as shown in Figure 3. The [100] in-plane crystallographic axis of the PMN-PT substrate is aligned with the sample's y-direction while the [011] in-plane crystallographic axis is aligned with the x-direction. The nanodot pitch along the y-direction is 500nm for all arrays while the nanodot pitch along the x-direction is 250nm and 150nm for arrays 1 and 2, respectively. Planar electrodes are first deposited on the PMN-PT's top (50 nm Pt) and bottom (50nm Au) surfaces. The nanostructure arrays are subsequently patterned on the Pt electrodes using a double layer of polymethylmethacrylate resist (PMMA 950K A2) and liftoff-assisting copolymer (EL6 MMA). Electron beam writing was performed using a charge dose of 700 C/cm$^2$. The resist pattern was developed using a solution of 1:3 MIBK to IPA (methyl isobutyl ketone and isopropyl alcohol, respectively). Prior to depositing the Ni nanostructures, the PMN-PT was poled with a 0.8 MV/m electric field. After poling, 5 nm Ti followed by 12 nm Ni (adhesion and ferromagnetic layers) were deposited by e-beam evaporation. This was followed by 12 hours of room-temperature lift-off using n-methyl-2-pyrrolidone (NMP) solvent.

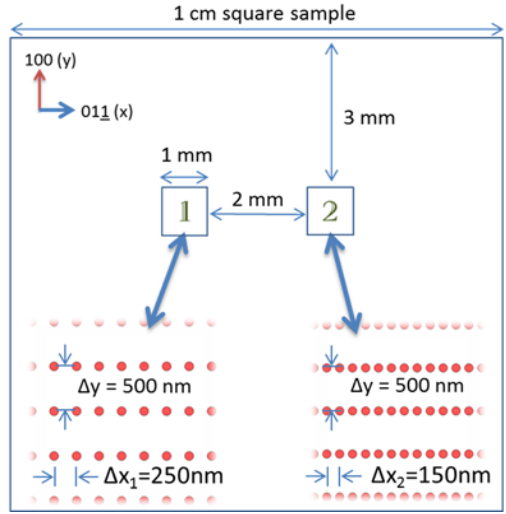

**Figure 3.** Schematic of tested sample with piezoelectric substrate crystallographic axes depicted.

Magnetization measurements are made using longitudinal MOKE, where the sample's plane is parallel to the applied field. The laser spot size for the MOKE measurement system is 2mm, hence measured M-H curves show the average array response for either arrays 1 or 2. The sample is held in place between the poles of the electromagnet using custom holders made of brass clips and PCB board. M-H curves are measured in both planar directions (i.e., x and y) and with an electric field of 0.6 MV/m applied.  An electric field of 0.6MV/m produces anisotropic in-plane strain[28]; specifically the strains are $\varepsilon_x = 1000\mu\epsilon$ and $\varepsilon_y = -3000\mu\epsilon$.  Each published M-H curve consists of up to 500 hysteresis loops sampled per orientation and electric field in order to reduce error.

Figure 4a shows M-H curves for the 250nm spacing measured along both x & y directions without an electric field. The M vs. H curves for both x and y directions are similar indicating an isotropic in-plane magnetization response. All the nanodots are circular, and thus shape does not introduce any magnetic anisotropy.  Furthermore, since the measurement shows an absence of in-plane magnetic anisotropy, one concludes that the spacing between nanodots is sufficiently large to eliminate any dipole coupling. Specifically, 250 nm spacing is sufficiently large that the dipole-dipole coupling range is trivial for these circular Ni nanodots. We calculate that the effective field due to this coupling is ~ 6 Oe even if both nearest neighbors are included.

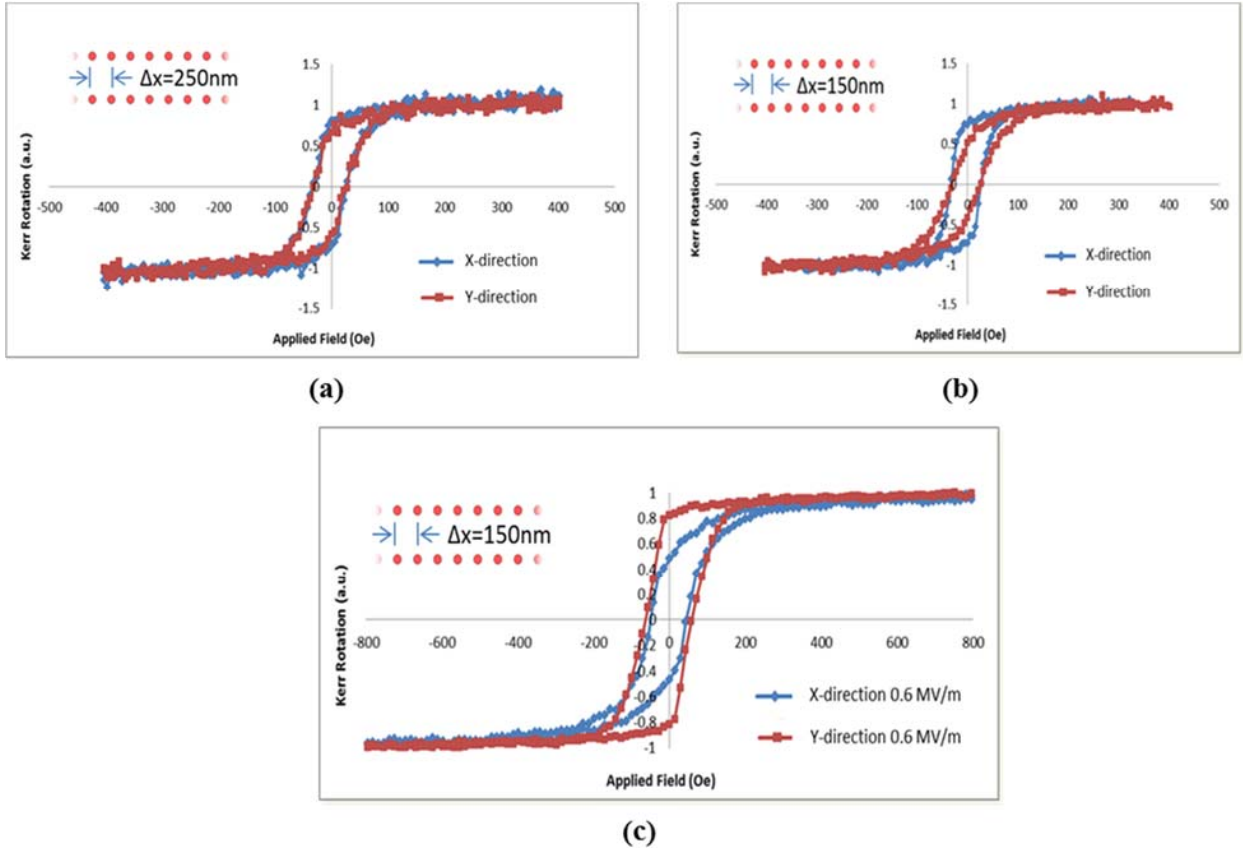

**Figure 4.** M-H curves for array configurations 1 and 2. (a) Magnetization data for configuration 1 (i.e., 250 nm spacing) at no electric field with applied magnetic field along both planar directions. (b) Magnetization data for configuration 2 (i.e., 150 nm spacing) at no electric field with applied magnetic field along both planar directions. (c) Magnetization data for configuration 2 at 0.6 MV/m electric field with applied magnetic field along both planar directions.

Figure 4b shows M-H loops measured along both x and y directions for the 150 nm spaced array without an electric field. When contrasting these measurements to those of Figure 4a for the 250 nm spaced array, distinct similarities and differences are noted. First, there is a similarity in the $M_r$ values measured in the x direction, in particular they are approximately 0.8. Second, there is a dissimilarity in the $M_r$ values measured in the y direction; specifically the $M_r$ for the 150 nm array is 0.5 while for the 250 nm array it is 0.8. The results for the 150 nm spacing clearly shows the presence of dipole-dipole coupling. Specifically dipole-dipole coupling along the y direction should produce anti-parallel magnetization alignment reducing the measured $M_r$ values along the y direction.

Figure 4c shows M-H curves for the 150 nm spaced array with a 0.6 MV/m electric field applied. The results show a distinct change in the M-H curves with the applied electric field as contrasted with the results without an electric field shown in Figure 4b. Specifically, without an electric field the remnant magnetization in the x-direction is 0.8, but when a field is applied the remanence decreases to approximately 0.5. The remanence decreases as the easy axis is now being established along an axis perpendicular to the x-axis array. This causes the easy magnetization direction to rotate toward the y-axis and away from x-axis, in spite of the dipole coupling. A change in $M_r$ occurs for the y-direction as well. However, the remanence increases from 0.5 to about 0.8 with an applied electric field. The remanence increase in the y-direction is surprising given the dipole coupling favors an anti-parallel state in the y direction. However, for this configuration the stress induced magnetic anisotropy is sufficiently large to mask the dipole effect and ensure the nanomagnets magnetization remains along the y-direction even after the magnetic field is withdrawn. This result clearly confirms that the magnetic easy axis of the nanomagnet is established along the y-direction when an electric field is applied. This change is attributed to the voltage induced magnetoelastic effect. Since Ni is negatively magnetostrictive, the magnetoelastic response of the nanostructures causes the magnetization to favor alignment with the axis of compression implying that the chain axis is magnetically harder. We also established that the magnetoelastic energy is sufficiently larger than the dipole-dipole coupling that favors anti-parallel alignment in the y-direction and parallel alignment in the x-direction

The simulations and the experiments together demonstrate the feasibility of using circular magnetostrictive nanomagnets clocked with strain to propagate information. We have further estimated theoretically that information can be transmitted along a circular nanomagnet chain clocked with stress with high reliability (error $< 10^{-6}$ m, see supplement section B) considering there are no defects than pin the magnetization in such magnets.  It can also be extremely low energy $< 1$aJ/bit (see Supplement, section C), provided it is implemented on thin film PZT. More importantly, this could provide a path to the ultimate scaling of nanomagnetic devices to implement Boolean operation and propagate logic at lateral dimensions below 20 nm, along with very little energy dissipation. This paradigm will be *both* scalable and extremely energy-efficient provided lithographic variations, misalignments and imperfections can be controlled at these scales.

ASSOCIATED CONTENT

**Supplementary Information**

This material is available free of charge via the Internet at http://pubs.acs.org.

AUTHOR INFORMATION

**Corresponding Author**

*jatulasimha@vcu.edu (J. A)

**Author Contributions**

All authors played a role in conception of the idea, planning the experiments, discussing the data, analysing the results and writing the manuscript. M.S.F and M.M.A performed the simulations; W.Y.S. and P.N. performed the experiments.

**Notes**

The authors declare no competing financial interests.

ACKNOWLEDGEMENTS

J. A., M.S.F and M.M.A acknowledges support from NSF CAREER grant CCF-1253370. W.Y.S., P.N, A.C.C. and G. P.C were supported in part by FAME, one of six centers of STARnet, a Semiconductor Research Corporation program sponsored by MARCO and DARPA.

# Binary information propagation in circular magnetic nanodot arrays using strain induced magnetic anisotropy

*M. Salehi-Fashami[1], M. Al-Rashid[2,3], Wei-Yang Sun[4], P. Nordeen[4], S. Bandyopadhyay[3], A.C. Chavez[4], G.P. Carman[4] and J. Atulasimha[2,3]\**

[1]Department of Physics, Univ. of Delaware, Newark, DE 19716.

[2]Department of Mechanical and Nuclear Engineering, Virginia Commonwealth Univ., Richmond, VA 23284.

[3]Department of Electrical and Computer Engineering, Virginia Commonwealth Univ., Richmond, VA 23284.

[4]Department of Mechanical and Aerospace Engineering, Univ. of California, Los Angeles, CA 90095.

\*Email: jatulasimha@vcu.edu

In the main paper, we demonstrated strain-clocked information propagation in an array of circular nanomagnets clocked with stress. In this supplement we provide more details about the simulations and experiments with equations, supplementary figures and discussion.

## SUPPLEMENTARY SECTION A: NUMERICAL SIMULATION

### A1. How uniaxial stress can induce bistability for the magnetization orientation in a circular nanomagnet

The potential energy landscape in the plane of a circular single-domain nanomagnet as a function of the magnetization orientation (angle $\phi$) is shown in Figure S1 in the absence and presence of stress. In the absence of stress, there is no energy minimum and hence no stable orientation of the magnetization vector. However, when compressive stress is applied to a material with negative magnetostriction (tensile stress to a material with positive magnetostriction), degenerate energy minima develop at $\phi$ = 90° and -90° (270°), if the stress is applied along $\phi$ = 90° (or -90°). Therefore, the magnetization tends to align along one of these two stable orientations and the magnetization direction becomes bistable. Thus, even an ultrasmall isotropic particle can be given stable magnetization orientations. This is the essence of making nanomagnetic computing technology, utilizing nanomagnets with bistable magnetization, *scalable*.

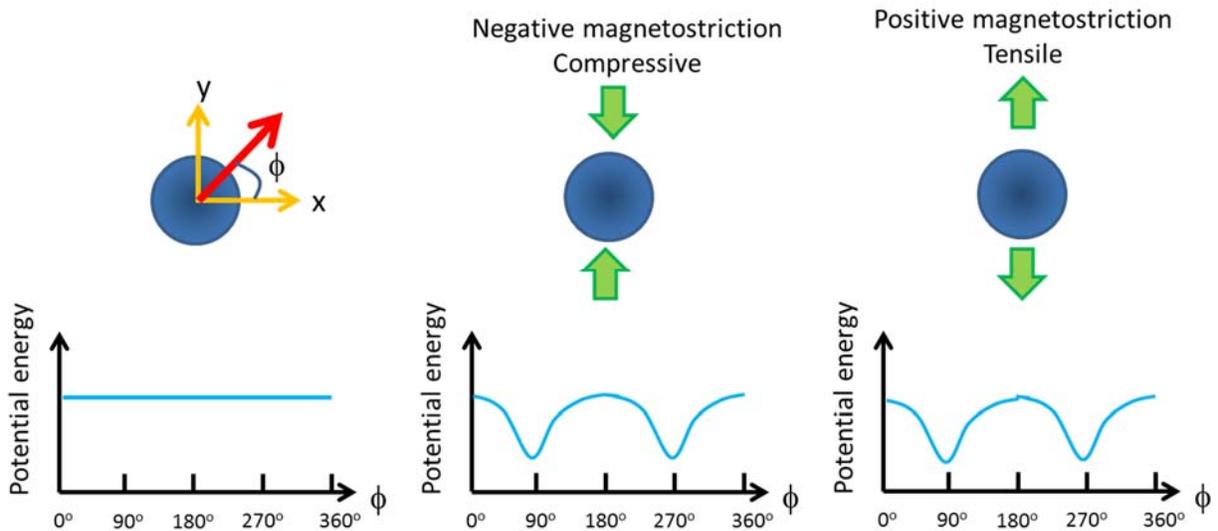

**Figure S1:** Energy profile of a circular nanomagnet as a function of in-plane azimuthal orientation of the magnetization at no stress (left), compressive stress (centre) and tensile stress (right).

**A2. Simulation of magnetization dynamics using Landau-Lifshitz-Gilbert equation in the presence of thermal noise.**

The stochastic Landau-Lifshitz-Gilbert (LLG) equation governing the magnetization dynamics in a single-domain nanomagnet is

$$\frac{d\vec{M}(t)}{dt} = -\gamma\vec{M}(t)\times\vec{H}_{eff}(t) - \frac{\alpha\gamma}{M_s}\Big[\vec{M}(t)\times\big(\vec{M}(t)\times\vec{H}_{eff}(t)\big)\Big] \qquad , \qquad (1)$$

with the effective field being given by

$$\vec{H}_{eff}^{\ i}(t) = -\frac{1}{\mu_0\Omega}\frac{\partial U_i(t)}{\partial\vec{M}_i(t)} = -\frac{1}{\mu_0 M_s\Omega}\nabla_{\vec{m}}U_i(t) + H_{thermal}(t) \qquad , \qquad (2)$$

where $\Omega$ is the volume of the nanomagnet, $H_{thermal}(t)$ is the magnetic field acting on the magnetization owing to thermal noise and $U_i(t)$ is its total potential energy at the instant of time $t$.

In an array of dipole-coupled closely spaced circular nanomagnets in the *x-y* plane, the total potential energy of each nanomagnet is:

$$U_i(t) = \sum_{\substack{i\neq j \\ i=j-1\ to\ j+1}} E_{dipole-dipole}^{i-j}(t) + \Omega\left(\frac{M_s^2\mu_o}{2}\right)\big[N_{d-xx}m_{x_i}^2(t) + N_{d-yy}m_{y_i}^2(t) + N_{d-zz}m_{z_i}^2(t)\big] -$$

$$\Omega\left(\frac{3}{2}\lambda_S\sigma_i(t)\right)m_{y_i}^2(t)\ ) \qquad , \qquad (3)$$

where $N_{d\_kk}$ is the demagnetization factor in the $k^{th}$ direction. The reduced magnetization can be defined as:

$$\vec{m} = \frac{\vec{M}}{M_S};\ m_x^2 + m_y^2 + m_z^2 = 1 \qquad\qquad (4)$$

In equation (3), the first term represents the dipole-dipole interaction energy ($E_{dipole-dipole}$) between nearest neighbor magnets. The second term, $E_{shape-anisotropy}$, denotes the shape anisotropy energy due to the circular cylinder shape (although $N_{d\_xx}=N_{d\_yy}$ because the magnets have circular cross-section, $N_{d\_zz}$ is different from $N_{d-xx}$ or $N_{d-yy}$ and the magnetization dynamics is significantly affected by out-of-plane excursions of the magnetization).The third term represents stress anisotropy energy caused by the stress ($\sigma$) that is developed in the magnetostrictive layer when strain is transferred to it from the bottom piezoelectric layer that is strained by the application of voltage as shown in Fig. 1 of the main paper.

The effect of thermal perturbation is modeled with a random field ($\vec{H}_{thermal}$) with statistical properties described below:

$$\vec{H}_{thermal}(t) = \sqrt{\frac{2K_B T \alpha}{\mu_0 M_s \gamma \Omega \Delta t}} (\vec{G}(t)) \tag{5}$$

where $\vec{G}(t)$ is a Gaussian random distribution with mean of 0 and variance of 1 in each Cartesian coordinate axis; $\Delta t = 1(ps)$ is the time step used in simulating the switching trajectories and it is inversely proportional to the attempt frequency with which thermal noise disrupts magnetization.

The stochastic LLG equation (Equation (1)) is solved ~100,000-1,000,000 times to obtain magnetization switching statistics under the influence of stress, dipole interaction and thermal noise. First, an equilibrium distribution for the initial magnetization orientation (in time) is generated by running the simulation to solve Equation (1) in the presence of only thermal noise (no stress or dipole interaction) for a very long time and sampling the magnetization orientations at regular periods. Next, stress and dipole interactions are turned on and Equation (1) is solved in the presence of thermal noise to generate $10^5 - 10^6$ switching trajectories. For generating each trajectory, the initial magnetization orientation is picked from the generated equilibrium distribution with appropriate statistical weight.

**A3. Calculating demagnetization factors of the circular magnetic dots:**

The demagnetization factors $N_{d-kk}$ for the circular cylinder nanomagnet used in this simulation in the x, y and z direction are denoted by $N_{xx}$, $N_{yy}$ and $N_{zz}$. In general

$$N_{xx} + N_{yy} + N_{zz} = 1 \qquad (6)$$

and specifically for the cylindrical geometry (circular dots)

$$N_x = N_y \qquad (7)$$

If the length of the cylinder is $2l$ and the radius is $a$, then the demagnetizing factor in the z direction can be expressed approximately as[1]

$$N_z = 1 - 2lL_s/(\mu_0 \pi a^2) \qquad , \qquad (8)$$

where

$L_s$ is the self-inductance given by [1]:

$$L_s = \frac{2\mu_0}{3l^2}[\sqrt{a^2 + l^2}\{l^2 F(k_s) + (a^2 - l^2)E(k_s)\} - a^3] \qquad (9)$$

In the above equation, $F(k_s)$ and $E(k_s)$ are the complete elliptical integrals of the first and second kind and $k_s$ is defined as:

$$k_s = \frac{a^2}{a^2 + l^2} \qquad (10)$$

**<u>Supplementary Section B: Reliability of information transmission in the presence of thermal noise</u>**

An important consideration in any logic scheme, is the reliability with which information is propagated in the presence of thermal noise. We simulated magnetization dynamics in the presence of thermal noise $10^5$ times at each pitch (center-center distance) and counted the number of times propagation fails. We model failure owing to disruption of the switching due to thermal noise and not owing to other potential spoilers such as pinning of magnetization by defects or misalignment (magnet centers are not on the same straight line). Fig S2 shows the switching probability of the chain of circular nanomagnets as a function of the center-center distance (pitch) between the 100 nm diameter and 12 nm thick circular nanomagnets clocked with a stain ~250 ε-strain. Clearly, an increase in pitch leads to lower dipole coupling, making it easier for thermal noise to cause disruptions in the magnetization switching process. This leads to decreased switching probability with increasing pitch. For one particular case, the design presented in this paper (pitch ~150 nm), there was no switching error in $10^6$ simulations indicating the error probability is smaller than $10^{-6}$.

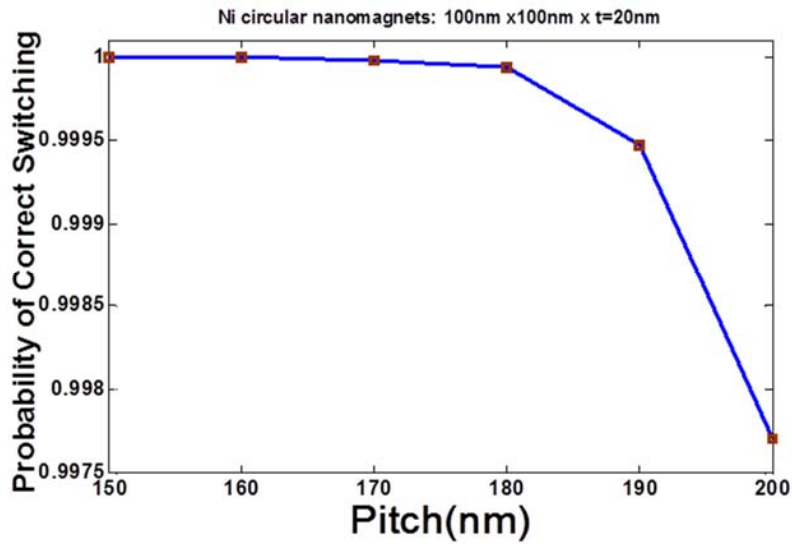

**Figure S2.** Probability of switching in circular nanomagnets for different dipole coupling under the effect of thermal noise. The coupling increases with decreasing pitch (center-center) separation resulting in increased switching reliability.

## Supplementary Section C

We calculate the energy dissipation in the clock (charging and discharging the piezoelectric layer under the two electrodes) per clock cycle, per nanomagnet. To generate a stress ~50 MPa in the Ni nanomagnet with Young's modulus ~200 GPa, we would need to transfer a strain of ~250 micro-strain. If a local strain-clocking scheme is employed, stress can be applied selectively to targeted nanomagnets in the manner described in of Fig 1 of main paper, Cui et al.[2] and Liang et al[3]. A conservative estimate of the electric field needed for PMN-PT with $d_{33}$ = ~(1500-2500) pm/V and $d_{31}$ = ~ -(700-1300) pm/V [see Ref 4 as well as http://www.trstechnologies.com/] in the above configuration is ~250 kV/m. To apply this field locally between the electrode and the substrate for a PMN-PT film of thickness t~100 nm, the voltage required would have been ~25 mV. The capacitance between the electrode pair and substrate is calculated by treating them as two flat plate capacitors in parallel. The area of each plate is A = $10^{-14}$ $m^2$ (assume square electrode of width ~100 nm). The total capacitance including both electrodes is, C = $2\varepsilon_0\varepsilon_r A/ t$ is ~5 fF. Assuming all the energy involved in charging the capacitor to strain the nanomagnet is lost, the energy dissipation/clock cycle, E= $(1/2)CV^2$ = $1.56\times10^{-18}$ J (1.56 aJ). Additional dissipation in the magnet due to Gilbert damping must then be taken into account but this ~1 aJ per clock cycle for a 1 GHz clock[5]. Scaling the nanomagnet dimensions further and using highly magnetostrictive materials such as Terfenol-D will drive the energy dissipation below ~1aJ.